\documentclass[11pt,twoside,onecolumn]{article}
\usepackage[]{epsfig}
\newcommand\<{{\langle}}
\renewcommand\>{{\rangle}}
\renewcommand\O{{O}}

\newcommand\U{{\cal U}}
\newcommand{\be}{\begin{eqnarray}}
\newcommand{\ee}{\end{eqnarray}}

\title{\bf Cosmological Unparticle Correlators}
\author{G.L. Alberghi$^{1.2}$\thanks{e-mail: alberghi@bo.infn.it},\
A.Yu. Kamenshchik$^{1.3}$\thanks{e-mail: kamenshchik@bo.infn.it},\
A. Tronconi$^{1}$\thanks{e-mail: tronconi@bo.infn.it},\
\\G.P. Vacca$^{1}$\thanks{e-mail: vacca@bo.infn.it}
\ and G. Venturi$^{1}$\thanks{e-mail: armitage@bo.infn.it}\\
 \\
$^{1}${\em Dipartimento di Fisica, Universit\`a di Bologna, and}
\\
{\em Istituto Nazionale di Fisica Nucleare,
Sezione di Bologna, Italy}\\
$^{2}${\em Dipartimento di Astronomia, Universit\`a di Bologna, Italy}
\\
$^{3}${\em L.D. Landau Institute for Theoretical Physics of Russian 
Academy of Sciences,} \\ {\em Kosygin street 2, 119334 Moscow, Russia.}} 
\begin{document}
%
%
\maketitle
\begin{abstract}
\noindent
We introduce and study an extension of the correlator of unparticle
matter operators in a cosmological environment. 
Starting from FRW spaces we specialize to a de Sitter spacetime and
derive its inflationary power spectrum which we find to be almost flat.
We finally investigate some consequences of requiring the existence of a
unitary boundary conformal field theory in the framework of the dS/CFT
correspondence.

\end{abstract}
%
\pagestyle{plain}
\raggedbottom
\setcounter{page}{1}
%
\section{Introduction}
Recently \cite{Georgi:2007ek}, Georgi has suggested that a scale-invariant
sector, arising from an unknown complicated non-linear theory with a
non-trivial infrared fixed point, may exist and be coupled to the Standard
Model (SM) via higher-dimensional operators suppressed by a large mass scale $M_{U}$. 
Due to scale-invariance, this sector in the low energy effective field theory
might not be described in terms of particles. Following
Ref.~\cite{Georgi:2007ek}, we will refer to this sector collectively as ``the
unparticle" sector. In this and subsequent works~\cite{Georgi:2007si,pheno}
the unusual properties associated with the scaling of the apparent phase space
volume, the missing energy spectra, interference patterns, etc., have been analyzed.
These properties can give origin to some interesting collider and flavor
phenomenology that may appear above the TeV scale or in precision measurements.

Lately some authors (see Refs. in \cite{Cosmo}) have also examined some
astrophysical and cosmological implications of the existence of such an
unparticle sector. They have dealt, for example, with bounds from
nucleosynthesis and stellar evolution and with neutrino astrophysics.
\par
In a cosmological framework it is natural to address the question whether such
an unparticle sector can contribute to cosmic dark matter density and which
signatures such a sector could present. The possibility that a deformed version
of such a phase exists is of course expected at least physically in a geometry
slightly deformed from the Minkowski case.

Assuming that this may be realized
more in general, we shall therefore consider unparticle matter in
Friedmann-Robertson-Walker (FRW) cosmological spaces and study in more detail the 
de~Sitter space-time case, which may provide a description of the phase with
an extremely small cosmological constant inferred by recent astronomical
measurements as well as an approximation  to the inflationary phase in the far past. 
The correlator of unparticle operators may be taken to describe a contribution
to the cosmological spectrum below a certain transmutational scale $\Lambda_{\cal{U}}$.
In particular we study the spectrum of correlators of unparticle matter
characterized by a  $\Lambda_{\cal{U}} \ll H$ as may be reasonably assumed in
the inflationary regime.   

Studying the unparticle phase in such de Sitter patches of our universe can be
constrained by requiring the support of the unparticle spectral function of
the operator of interest to be compatible with the existence of a dual unitary
conformal field theory on its boundary, within the framework of the dS/CFT
correspondence \cite{Strominger}.
If taken seriously this constraint roughly leads to requiring a
transmutational scale, associated with the effective low energy unparticle
theory, to be not greater than the cosmological Hubble parameter. 

\section{Unparticle Decomposition in Flat Space}
\setcounter{equation}{0}
\label{Decomposition}
Let us briefly review some facts related to a possible unparticle sector in Minkowski space.
Following Ref.\cite{Georgi:2007ek}, let us imagine that there exists a
scale invariant sector of our world described, e.g., by some strongly
self-coupled conformal theory, weakly interacting with the Standard Model
sector. Such an unparticle sector might be generated, for example, by
non-perturbative effects in a conformally-invariant Banks-Zaks (BZ) sector,
consisting of vector-like non-Abelian gauge fields and massless fermions \cite{bz}.  
The BZ sector undergoes dimensional transmutation to the unparticle phase
below the energy scale $\Lambda_{\cal U}$ and its low energy effective description
is associated with a CFT with a non-trivial infrared fixed point
(i.e. such a possibility excludes explicitly the case of a trivial gaussian IR fixed point).
This unparticle phase has no particle description but interacts with the
SM via unparticle operators $O$ of mass dimension $d_{\cal{U}}$. 
The couplings and the dimensions of the unparticle operators will depend
on the underlying BZ operators and the infra-red dynamics of dimensional transmutation.
Since the unparticle sector is selfinteracting, the dimension of $O$ can be
nontrivial (noninteger).  The correlation function of such an operator in
flat spacetime can be written as~\cite{Georgi:2007si,Stephanov:2007ry}
\begin{equation}
  \label{eq:OO}
  \int d^4x e^{iPx}\<0|T\O(x)\O^\dag(0)|0\> \\
= \int \frac{dM^2}{2\pi} \rho_\O(M^2) \frac{i}{P^2-M^2+i\varepsilon}\,.
\end{equation}
By scale invariance the spectral function of the operator~$\O$ must
be a power of $M^2$:
\begin{equation}
  \label{eq:rhoO-scaling}
  \rho_\O(M^2) = A_{d_\U} (M^2)^{d_\U-2},
\end{equation}
where $A_{d_\U}$ is a normalization constant, chosen by convention.
Such a representation has been shown to emerge in the framework of
``unparticle deconstruction'' from towers of massive fields in the limit
of a vanishing mass spacing parameter~\cite{Stephanov:2007ry}.

\section{The Unparticle Correlators in FRW and de Sitter spaces}
One of the questions we wish to address is what can one expect when
considering cosmological and, in particular, de Sitter instead of flat
spacetimes. Indeed, in the process of investigating all possible physics one
may find in our universe, one is completely justified in also imagining the
occurrence of an effective unparticle phase at low energies in spaces close to
Minkowski described, for example, by FRW and de Sitter geometries.
We also consider the hypothesis that in such spaces, even for cases far away from
Minkowski such as the inflationary regime, a deformed unparticle phase might
exist. In our analysis we restrict ourself to flat spatial sections.
We feel that the problem of demonstrating the existence of non trivial phases
in the low energy regime of interacting quantum fields theories on curved
spaces deserves proper investigation. In the following, assuming this
possibility,  we shall restrict ourself to analyzing some consequences
of its presence.  
 
Again a possible partial description of an unparticle phase may be
given in terms of the properties of correlators of unparticle operators.
In particular we shall consider an extension to such spaces of the two point correlators,
extension directly related to the fact that the Green's funtions and the
field Fourier modes of the fields in curved space time reduce to
the flat space ones when the curved geometry approaches the Minkowski one.

Let us begin by considering a (conformally) flat FRW spacetime with metric
\be
ds^2=-dt^2+a^2(t) d{\bf x}^2= a^2(\eta)\left(-d\eta^2+ d {\bf x}^2\right)\,.
\ee
On using a decomposition similar to eq. (\ref{eq:OO}) we consider the
contributions of any massive mode of a scalar field (deconstructed
representation) which sum up to the unparticle phase after integration
with the corresponding spectral function $\rho_O(m^2)$, which we assume
unchanged for the chosen gravitational regime, although in general it cound
depend on the Hubble scale.
In particular we write a spectral representation for the correlators of the
hermitian unparticle operators $\O$,
$\<0|\O(\eta,{\bf x})\O(\eta',{\bf x}')|0\>$, in momentum space:
\be
  \int d^3{\bf x} e^{i {\, \bf k \, x}}
\<0|\O(\eta, {\bf x})\O(\eta',{\bf 0})|0\> 
= \int \frac{dM^2}{2\pi} \rho_\O(M^2)
\int {d^3{\bf x} \over (2\pi)^3} e^{i {\, \bf k \, x}}
\<0|\varphi_M(\eta, {\bf x}) \varphi_M(\eta',{\bf 0})|0\> \,,
\ee
where we represent unparticle matter in a form deconstructed in terms of the
continuous set, for simplicity, of minimally coupled scalar fields $\varphi_M$
and the unparticle spectral function $\rho_O$.

Clearly for any explicit evaluation the choice of vacuum is crucial.
Indeed in order to quantize the quantum scalar fields on the FRW curved
space~\cite{BD} one must choose the set of creation and annihilation operators
for any mode expansion of the field
$\varphi(\eta,{\bf x})=\sum_k [a_k e^{i {\, \bf k \, x}} \varphi_k(\eta) +
a_k^\dag  e^{-i {\, \bf k \, x}}\varphi^*_k(\eta)]$ adopted.

On using standard Fourier modes and considering the coincident time correlators we define
\be
    P_{\cal{U}}(k, \eta) = 
 \int_0 ^{\infty} P_{\varphi}(k, \eta, m) \rho_O (m^2) d  m^2,
\label{PU}
\ee
where the power spectrum $P_{\varphi}$ for a scalar field in FRW is
obtained through the relation 
\be
\langle \hat \varphi (\eta, {\bf x}) \hat \varphi (\eta, {\bf x'})
\rangle = \int_0^{+\infty} \frac{dk}{k} 
\frac{\sin k|{\bf x} - {\bf x'}|}{k \, |{\bf x} - {\bf x'}|}
\frac{k^3 | \varphi_k (\eta)|^2}{2 \pi^2}
\equiv \int_0^{+\infty} \frac{dk}{k} 
\frac{\sin k|{\bf x} - {\bf x'}|}{k \, |{\bf x} - {\bf x'}|}
P_\varphi (k,\eta,m) \,.
\label{twopoint}
\ee
We obtain
\be
    P^{FRW}_{\varphi}(k, \eta,m) = {1 \over 2 \pi^2} k^3 | \varphi_k (\eta)|^2\,,
\ee
where $\eta$ is the conformal time and $\varphi_k (\eta) = \chi_k(\eta)/a(\eta)$ is
written in terms of a solution of the differential equation
\be
{d^2\chi_k \over d \eta^2}+ \left(k^2 +m^2 a^2(\eta) -
{a''(\eta) \over a(\eta)}\right)\chi_k =0\,,
\ee
to which one must add an appropriate boundary condition.
For a generically given $a(\eta)$ no analytic solution for massive scalar
modes are available. Clearly the Minkowski limit is recovered for $a(\eta) \to 1$. 

De Sitter space with $a(\eta)=-1/(H \eta)$ for $-\infty<\eta<0$ is a special
case. Indeed in a spacetime described by the de Sitter geometry and
on choosing a Bunch-Davies vacuum one finds
\be
    P^{dS}_{\varphi}(k, \eta,m) = {H^2 \over 8 \pi} z^3 H^{(2)} _{\nu} (z)
    H^{(1)} _{\nu ^{*}} (z)
\label{pow_scal_dS}
\ee
with $ \nu^2 = 9  /4 - m^2 / H^2 $ and $ z = k / a H$. Again one can check,
using the asymptotic expansion of the Bessel functions appropiately continued
to an imaginary large index, the expected behavior in the Minkowski limit for
the Fourier modes
\be
\phi_{\bf k}(t,{\bf x})\sim e^{i \,{\bf k}\,{\bf x}} e^{-{3 \over 2} H t}
\frac{1}{\sqrt{H}}  H^{(2)} _{\nu} \left({k E^{-H t} \over H}\right)
 \stackrel { H \to 0}{\longrightarrow} \, \,
\frac{1}{(m^2+k^2)^{1/4}} e^{i \,{\bf k}\,{\bf x}+i \sqrt{m^2+k^2} t} \,,
\ee
where we omit an irrelevant proportionality factor not dependent on the coordinates
resulting from the non commutativity of the $t\to -\infty$ (equivalent
to $k \to \infty$) and $H \to 0$ limits.
 
In coordinate space, for different times, one has for the massive scalar
correlator the corresponding form, which is a direct solution of the
Klein-Gordon equation in de Sitter space:
\be
\langle \hat \varphi (\eta, {\bf x}) \hat \varphi (\eta', {\bf x'})
\rangle_{\rm dS} = \frac{H^2 \, \sec (\pi \nu)}{16 \pi^2}
\left(\frac{1}{4} - \nu^2\right) \, _2F_1 \left[ \frac{3}{2} +
\nu, \frac{3}{2} - \nu \, ; 2 \, ; 1
+\frac{(\eta-\eta')^2- | {\bf x} - {\bf x'}|^2}{4 \eta \eta'} \right] \,.
\label{twopoint_dS}
\ee

The explicit expression for the spectral function depends on the kind of
theory which undergoes dimensional transmutation towards the infrared region,
once the energy scale $\Lambda_{\cal U}$ is reached,
and can be postulated to be
\be
     \rho_O(m^2) = A_d (m^2)^{d-2} = A_d (m^2) ^{\Delta -1} \quad , \,m^2 <
     \Lambda_{\cal U}
\label{spectral_function}
\ee
where, following \cite{Georgi:2007ek,Georgi:2007si}, $ 0 < \Delta <1 $.

On retaining the assumptions valid for the inflationary regime,
characterized by large energy scales, the Hubble parameter being typically
$H \sim 10^{-5} M_{pl}$, we feel it quite reasonable to consider an
effective unparticle phase which takes place at $\Lambda_{\cal U} \ll H$.
On inserting such a relation together with the one in (\ref{pow_scal_dS}) into
eq.~({\ref{PU}) the spectral function may be computed:

\be
    P_O^{dS}(k,\eta) = \int_0 ^{\Lambda_{\cal U}^2} 
    P^{dS}_{\varphi}(k,\eta, m) \rho (m^2) d  m^2\simeq 
A_d \int_0 ^{\Lambda_{\cal U}^2} dm^2 (m^2) ^{\Delta -1}
{H^2 \over 8 \pi} z^3 {4^\nu z^{-2\nu}
    \Gamma[\nu]^2 \over \pi^2}\,.
\label{powdesitter}
\ee
The unparticle spectral function has a singular (but integrable)
behavior for small $m^2$. Therefore one notes that the unparticle correlator
is dominated by the integration around the origin and should not be very
sensitive with respect to the upper bound of the support of integration.
In this regime one may safely consider the approximation $\nu\simeq \frac{3}{2}
- \frac{x}{3}$ where we define $x=\frac{m^2}{H^2}$ and retain the lowest order
approximation for the expansion $\Gamma[3/2-x/3]\simeq \sqrt{\pi}/2$ and
keep the $x$ dependence only in the exponent. One finds:
\be
P_O^{dS}(k,\eta)\simeq \frac{A_d}{4}\left({3H \over 2}\right)^\Delta
    \left[\Gamma(\Delta)-\Gamma\left(\Delta,-\frac{2}{3} {\Lambda_{\cal U}^2 \over
     H^2} \ln{2\over z}\right)\right] \left(\ln{2\over z}\right)^\Delta \,,
\ee
where $ \Gamma (.,.) $ is the incomplete Gamma function.
The logarithm  of the power spectrum of eq. (\ref{powdesitter}) can be also
approximated to second order in $\Lambda_{\cal U}/H$ as:
\be
   \ln{P_O^{dS}(k,\eta)}\simeq  \ln{A_d
\left(\Lambda_{\cal U}^2\right)^{\Delta} \over 4\pi^2
     \Delta} +{2 \over 3} {\Delta \over 1+\Delta} \left({\Lambda_{\cal U}^2 \over
     H^2}\right) \left( \ln{k \over 2 a(\eta) H} +(2\ln 2+\gamma_E-2)\right)\,,
\label{logP}
\ee
where, in order to give a meaning to the logarithm of the power spectrum,
one should choose some unit for the energy scale.
Such a linear behavior with a small slope is completely confirmed by numerical investigations.
One therefore finds an almost gaussian spectrum with deviations (positive as
$\ln k$ increases) proportional
to $\Lambda^2_{\cal U}/H^2$, the logarithm of the momentum, and a
simple function of the anomalous dimension.
The entity of the correction increases as $ \Delta $ increases from 0 to 1.

One may be tempted to try to see if there is consistency with CMB data thus
eventually constraining the scaling dimension $\Delta$ and the energy scale
$\Lambda_{\cal U}$. Several issues should then be addressed.
In this analysis we have restricted ourselves to a de Sitter-like
inflationary expansion (constant Hubble parameter $H$), which gives for a
massive scalar field as well as for a deconstructed unparticle phase a blue tilted
spectrum~\cite{FMVV}. In the presence of inflation with a
time varing $H$ further investigations would be needed and red tilted
contributions may appear.
Moreover the last WMAP three year data~\cite{Spergel:2006hy} show
compatibility with a running spectral index,
and its central value is red tilted if defined at high momenta
and blue tilted when defined at low momenta.
One should note that the contribution of an unparticle phase 
may affect only a portion of the observed spectrum
(depending on the transmutational scale $\Lambda_{\cal U}$).

Since we are unable to estimate the relative weight of unparticle matter on the scalar
curvature perturbations we just observe what would happen if its contribution is of the
same order of magnitude as the standard contributions to the spectrum. Considering the result in
eq. (\ref{logP}) the unparticle phase would give a higher blue shift
when the factor $\Delta  \left({\Lambda_{\cal U}^2 / H^2}\right)$ is not much
smaller than $1$. With $\Lambda_{\cal U} / H \sim 0.1$ and a maximal
$\Delta \sim {\cal O}(1)$ one has a spectral index shifted from $1$ by less
then $1\%$, which is well below the experimental errors.
The only conclusion one may draw is that CMB data do not exclude the
possibility that an unparticle phase exists for a very wide range of its
parameters.
 
\section{De Sitter space and implications from dS/CFT}
The scalar representations of the de Sitter group $SO(1,d+1)$ split
into three series\cite{Vilenkin,Takook}: the principal series for
$m^2  \ge (\frac{d}{2}H)^2$, the complementary series for 
$0 < m^2 < (\frac{d}{2}H)^2$  and the discrete series, whose only physical
interesting case is $m^2 = 0$. 
Recently Strominger \cite{Strominger} has given arguments in favor of the
hypothesis that a scalar field of mass $0 < m^2 < (\frac{d}{2}H)^2$ in
$dS^{d+1}$ can be related to a conformal field theory living on an $S^d$,
which may be identified as the boundary (timelike infinity) of
de Sitter spacetime\cite{Bros}. 
This proposal was formulated in analogy with the AdS/CFT correspondence
and arguments suggesting its existence were given in \cite{Witten}.
It is significant that it does not seem possible to formulate a unitary
CFT for massive fields with $m^2 \ge (\frac{d}{2}H)^2$.
In fact for $m^2 > 9 H^2 /4$, the conformal weights of the operator
$ \hat O _{\varphi} $ corresponding to the massive scalar field $ \varphi $ become complex.
This means that the boundary CFT is not unitary if there are stable scalars
with masses above this bound. 
On the other hand one may observe that this point of view is consistent
with the statement of \cite{Polyakov} that if de Sitter space is unstable,
it must be described by a non-unitary CFT.
We shall therefore consider this fact when trying to give a description
of the unparticle phase from the deconstructing point of view, which
assumes that such a continuous set of massive scalars contributes in
its entirety to the unparticle conformal phase. 

One can also try to consider a general flat FRW geometry, with the
dynamics encoded in the scale parameter $a(t)$.
It was conjectured by Strominger~\cite{Strominger2} that bulk time
evolution might be dual to the RG flow between different conformally
invariant fixed points, with the Hubble parameter possibly related
to the number of degrees of freedom
(a discussion related to Zamolodchikov's c-theorem).
The picture given is, for example, the flow on the boundary from
UV to IR corresponding in the bulk to the interpolation from the
inflationary phase in the past to the present de Sitter-like slow expansion phase. 
In this duality, extrapolated beyond the conjectured dS/CFT,
the discussion of the unitarity of the non-conformal theory on
the boundary is a completely open problem, since one only knows the
constraints from the fixed conformal points along the RG flow.

The simplest case we can analyze is characterized by a low energy scale
$\Lambda_{\cal U}\sim H$ so that following the considerations of 
\cite{Tolley:2001gg,Takook,Strominger} previously described in section $3$, 
we assume that the density $ \rho (m^2) $ is non-vanishing only 
for the finite interval $ 0 < m^2 < 9 H^2 / 4 $ (implying that we
consider only the complementary series and thus a stable de Sitter
space). With such a choice we have no scale other than $H$ in the system since we write
\be
    P_O^{dS}(k,\eta) = \int_0 ^{{9 \over 4} H^2} 
    P^{dS}_{\varphi}(k,\eta, m) \rho (m^2) d  m^2\,.
\label{PSdS_CFT}
\ee
The integral cannot be determined in closed form, thus we evaluate
it numerically.
The power spectrum is shown in figs $1$ and $2$ for two different values of
the anomalous dimension $\Delta$ as a function of the logarithm of $H/k$.
\par
We are interested in values of $k$ in the spectral region such that $z=k/aH
\ll 1$. Moreover the unparticle spectral function has a singular
(but integrable) behavior for small $m^2$, as was already observed in the
previous section.
Therefore one notes that the unparticle correlator is dominated by the
integration around the origin and should not be very sensitive with respect 
to the upper bound of the support of integration. 
We observe that the dependence on $k$ increases for higher values of the
anomalous dimension $\Delta$ associated with the scaling properties of
the unparticle operator under cosideration.
Therefore the behavior is closer to a constant power spectrum (gaussian) for a
parameter $\Delta$ close to zero.
\begin{figure}[t!]
\centering
\epsfxsize2.5in
\epsfbox{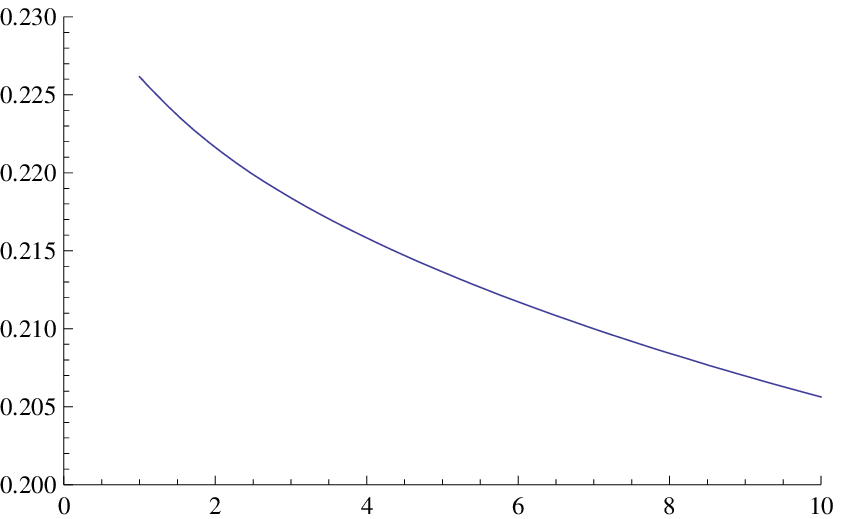} \hspace{1cm}
\epsfxsize2.7in
\epsfbox{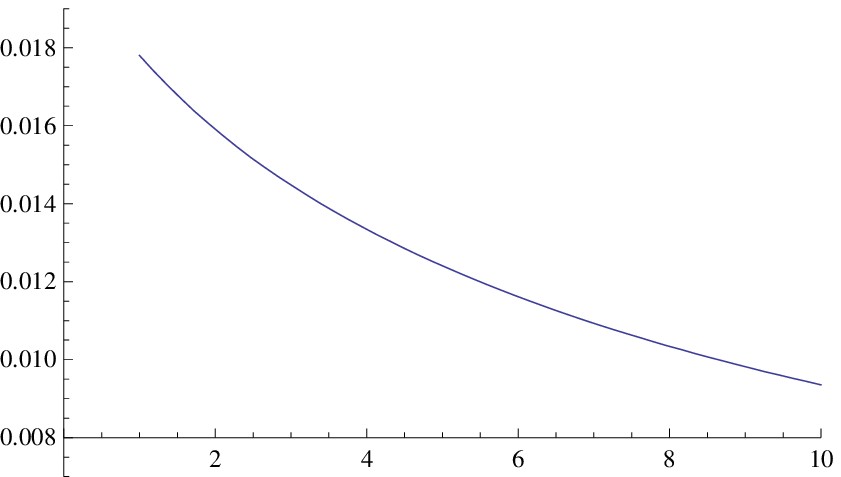}
\caption{\it De Sitter: Power spectrum (\ref{PSdS_CFT}) as function of $Log_{10}(H/k)$
for $\Delta = 1/10$ (left) and for
    $\Delta = 3/4$ (right).}
\end{figure}
%
\section{Conclusions}
We have addressed the question of describing an unparticle phase in a
cosmological scenario consisting of a FRW, in particular de Sitter, geometry.
In particular for an inflationary de Sitter patch of our universe we have
considered the two point function of unparticle operators and
its associated power spectrum. We have determined its analytical form
under the assumption of a transmutational scale small compared to the Hubble
scale. The resulting spectrum is almost gaussian with small deviations
regulated by the ratio of the cutoff with respect to the Hubble scales.
We observe that the dependence on $k$ increases for higher values of the
anomalous dimension $\Delta$ associated with the scaling properties of
the unparticle operator under cosideration.

In a stable de Sitter spacetime we have also tried to
deal with the consequences of assuming the existence of a dual unitary
conformal field theory, within the dS/CFT framework.
This fact seems to give constraints on the transmutational scale and
the anomalous dimension of the unparticle matter may strongly
affect the behavior of its correlators and power spectrum, with a 
behavior close to gaussianity for very small anomalous dimensions.
Also a smaller support of the spectral function due to a smaller
low energy transmutation scale is associated with a more gaussian spectrum.

We have only considered the case of a rigid space-time background.
In this context one may also study cases where the expansion is
associated to a time dependent Hubble parameter such as the
case of chaotic inflation which is characterized by a nearly
constant $\dot{H}$. Further the inclusion of fluctuations of the
metric and the associated issue of gauge freedom may of course
be considered. Such degrees of freedom may or may not be taken to
mix with the unparticle sector, leading to scenarios which should
nevertheless be described by some specific cases within the FRW framework.   

We note that if the low energy tail of the fluctuations of
a strongly interacting ``inflaton'' system could be described by an unparticle phase
(perhaps after mixing with fluctuations of the metric) the correlation of
curvature perturbations of the energy density would then be given by the result
of section $4$.
Of course a de Sitter geometry would be a very crude approximation
and would have to be replaced by a better description in terms of some FRW
cosmology.
Lastly one should ask oneself what the effective density-pressure equation of
state is for an unparticle phase and whether it is related to dark energy. 
\label{conc}
%
%
%
%

\end{document}